\documentclass[a4paper,12pt]{article}

\usepackage[dvipdfmx]{graphicx}
\usepackage{tikz}
\usetikzlibrary{decorations.pathmorphing}

\usepackage{amsmath}

\usepackage{mathrsfs}

\usepackage{bm}
\usepackage{color}
\usepackage{ulem}
\usepackage{amssymb}

\DeclareMathOperator{\Tr}{Tr}
\DeclareMathOperator{\PE}{PE}
\def\tr{\mathrm{tr}}
\def\Tr{\mathrm{Tr}}

\def\half{{1\over2}}

\newcommand{\wt}{\widetilde}

\def\={\stackrel{\bullet}{=}}

\def\({\left(}
\def\){\right)}
\def\[{\left[}
\def\]{\right]}

\def\cL{{\cal L}}

\def\cN{{\cal N}}

\def\cS{{\cal S}}

\def\mbf{\mathbf}

\def \be {\begin{equation}}
\def \ee {\end{equation}}
\def \beal#1 {\begin{align}#1\end{align}}
\def \bes#1 {\begin{equation}\begin{split}#1\end{split}\end{equation}}
\def \nn {\notag\\}

\def\sla#1{\not\!\!#1}

\usepackage{geometry}
\numberwithin{equation}{section}
\allowdisplaybreaks

\begin{document}

\newcommand{\hiduke}[1]{\hspace{\fill}{\small [{#1}]}}
\newcommand{\aff}[1]{${}^{#1}$}
\renewcommand{\thefootnote}{\fnsymbol{footnote}}

\begin{titlepage}
\begin{flushright}
{\footnotesize KIAS-P18033}, {\footnotesize YITP-18-29}
\end{flushright}
\begin{center}
{\Large\bf
% Comment on level/rank duality of minimal ${\cal N}=4$ Chern-Simons-matter theory
Index and duality of minimal ${\cal N}=4$ Chern-Simons-matter theories
}\\
\bigskip\bigskip
{\large Tomoki Nosaka\footnote{\tt nosaka@yukawa.kyoto-u.ac.jp}}\aff{1}
{\large and Shuichi Yokoyama\footnote{\tt shuichi.yokoyama@yukawa.kyoto-u.ac.jp}}\aff{2}\\
\bigskip\bigskip
\aff{1}: {\small
\it School of Physics, Korea Institute for Advanced Study\\
85 Hoegiro Dongdaemun-gu, Seoul 02455, Republic of Korea
}\\
\bigskip
\aff{2}: {\small
\it Yukawa Institute for Theoretical Physics, Kyoto University\\
Kitashirakawa-Oiwakecho, Sakyo, Kyoto 606-8502, Japan
}
\end{center}

\begin{abstract}

We perform a first step analysis toward generalization of the classification of $\cN=4$ linear quiver gauge theories by Gaiotto and Witten including Chern-Simons interaction. 
For this we investigate minimal ${\cal N}=4$ $\text{U}(N_1)_k\times \text{U}(N_2)_{-k}$ Chern-Simons theories and their superconformal indices. 
In the previous publication %arXiv:1706.07234 
we analyzed the three-sphere partition function of the theories, which implies that the theory is good/ugly/bad if $k-N_1-N_2$ is greater than/equal to/smaller than $-1$. 
In this paper we verify that this classification is consistent with the behavior of the superconformal index. 
We compare the superconformal indices for several pairs of non-bad theories connected by the Hanany-Witten transition and confirm their coincidence up to the contribution of one hypermultiplet.

\end{abstract}
\end{titlepage}

\renewcommand{\thefootnote}{\arabic{footnote}}
\setcounter{footnote}{0}

\tableofcontents

\section{Introduction and Summary}
\label{introduction}

Three dimensional gauge theories, in contrast to four dimensional ones, are always asymptotic free and strongly coupled in the infra-red (IR) regime. 
It is a central problem to investigate properties of their IR fixed points, and for this analysis instanton or monopole operators often play a key role. 

Indeed, (BPS) monopole operators played a crucial role in classifying a class of $\cN=4$ supersymmetric linear quiver gauge theories from their long distance behavior \cite{Gaiotto:2008ak}.
In this class it was already studied how to compute the R-charge of an arbitrary monopole operator \cite{Borokhov:2002ib,Borokhov:2002cg}, which depends on the number of hypermultiplets charged under the associated gauge group. 
If the number of hypermultiplets is great enough so that the R-charge of any monopole operator is always greater than half, then the system flows to a standard critical point.
In this case the theory was called ``good''.
On the other hand, if the number of hypermultiplets decreases to saturate some bound so that there exists a monopole operator whose R-charge is equal to half, then it decouples at the IR fixed point.  
This was called an ``ugly'' theory.
Finally if the number of hypermultiplets is so small that there exists a BPS monopole operator whose R-charge is less than half, then the IR fixed point of the system is not a standard one. This case was called a ``bad'' theory. 
We refer to such monopole operators in the ugly/bad theory as ugly/bad ones.
In a bad theory the R-symmetry in the IR has to be different from that in the UV in order for the system to preserve the unitarity.  

The infra-red physics of a bad quiver as well as the fate of such a bad monopole operator in the IR were further studied in \cite{Kapustin:2009kz,Bashkirov:2013dda,Yaakov:2013fza,Hwang:2015wna} (See \cite{Assel:2017jgo,Dey:2017fqs} for recent study).
The simplest example to study a bad theory is ${\cal N}=4$ SQCD with $U(N_\text{c})$ gauge group and $N_\text{f}$ fundamental hypermultiplets.\footnote{ 
In this case the theory is good if $2N_c\le N_f$, ugly if $N_f=2N_c-1$ and bad if $N_c < N_f\le 2N_c -2$ \cite{Gaiotto:2008ak}.} 
It was pointed out in \cite{Kapustin:2009kz} that this classification by using a monopole operator is related to the convergence property of the three-sphere partition function, which was shown to reduce to a finite dimensional integration by the supersymmetric localization, so that the partition function of a good/bad theory is absolutely convergent/divergent. 
The three-sphere partition function in a bad case was further studied by converting into a good or ugly theory by adding extra hypermultiplets with real mass so as to flow to the original bad theory in the IR \cite{Yaakov:2013fza}. 
This modified partition function was used to propose a Seiberg-like duality between a good theory and a bad one in $\cN=4$ SQCD, where the bad monopole operators decouple in the IR. 
This duality was further supported by a direct comparison of the factorized form of the superconformal index for both sides \cite{Hwang:2015wna}.\footnote{
Recent argument by \cite{Assel:2017jgo} based on the analysis of the moduli space suggests that the duality holds only around the special singular locus called ``symmetric vacuum'' in the moduli space of the bad theory.
This is, however, still consistent with the observations in \cite{Yaakov:2013fza,Hwang:2015wna}.
}

It is natural to ask whether this classification of $\cN=4$ supersymmetric linear quiver gauge theories works also for the theories including Chern-Simons interaction. 
This question may not be trivial in the following sense. 
First the three-sphere partition function for a supersymmetric Chern-Simons theory has the Fresnel factor, which plays a role of a damping factor by performing the analytic continuation for the Chern-Simons coupling, so that there may be no issue on the convergence or at least completely different convergence property from the case without Chern-Simons terms. 
This suggests that the classification by using the convergence property of the three-sphere partition function does not work or at least needs speculation. 
On the other hand, the classification by employing monopole operators seems to work, though the modification needs to be done once Chern-Simons interaction is introduced. 
This is because the gauge charge of bare monopole operators changes accordingly, and so does the total R-charge of the gauge invariant monopole operator due to the dress by compensating matter operators. 

The above question is strategically good to be addressed first by using the simplest model in the $\cN=4$ linear quiver Chern-Simons theories, that is ${\cal N}=4$ $U(N_1)_k\times U(N_2)_{-k}$ Chern-Simons theory coupled with single bi-fundamental hypermultiplet.
In our previous publication we studied the three-sphere partition function in this minimal $\cN=4$ Chern-Simons theory, where we performed the remaining integrations in the localization formula explicitly \cite{Nosaka:2017ohr}.
As a result we found that the resulting partition function diverges for $k -N_1 -N_2 \le -2$. This divergence is rather the IR one than the UV one since it can be cured by introducing FI terms or equivalently mass terms. 
We have further found that the resulting partition functions of the theory connected by the Hanany-Witten transition coincide up to an overall factor, which is given by a trigonometric function to the power of $k-N_1-N_2$.
These results are reminiscent of those for $\cN=4$ SQCD in \cite{Yaakov:2013fza} and suggests that the minimal $\cN=4$ Chern-Simons theory is bad for $k -N_1 -N_2 \le -2$ in the terminology of \cite{Gaiotto:2008ak}.
In this paper we study this behavior of minimal $\cN=4$ Chern-Simons theories by computing their superconformal indices \cite{Kim:2009wb,Imamura:2011su} (see \cite{Yokoyama:2011qu} for a review and \cite{Aharony:2015pla} for related works). 

The rest of this paper is organized as follows.
In section \ref{N4theory} we review the results of \cite{Nosaka:2017ohr} and compare them with the observations in \cite{Kapustin:2009kz,Yaakov:2013fza} without Chern-Simons terms.
We provide additional evidence for this duality by counting the dimension of the moduli spaces.
In section \ref{generalexpression} we introduce the superconformal index, and in section \ref{perturbative} we estimate the contributions from dressed monopole operator.
As a result we find that for $k-N_1-N_2\le -2$ there is a family of infinitely many monopole operator which contribute with non-positive power of $x$, the fugacity of $D+J_3$.
This implies that $U(N_1)_k\times U(N_2)_{-k}$ theory with $k-N_1-N_2\le -2$ is indeed a bad theory.
In section \ref{perturbative_results} we further compute the superconformal index of non-bad theories with $k-N_1-N_2=\pm 1$ explicitly in small $x$ expansion.
By comparing those for pairs of theories related through the Hanany-Witten transition, we find that the ratio of the superconformal indices have a completely same expression as the superconformal index of a hypermultiplet.
This suggests that the theory with $k-N_1-N_2=-1$ is ugly and dual to the paired good theory with $k-N_1-N_2=1$ plus a hypermultiplet.

\section{Minimal $\cN=4$ Chern-Simons matter theories }
\label{N4theory}

In this section we briefly review the minimal $\cN=4$ Chern-Simons matter theory or $\text{U}(N_1)_k\times \text{U}(N_2)_{-k}$ Chern-Simons theory interacting with one bifundamental hypermultiplet \cite{Gaiotto:2008sd}. (See also \cite{Hosomichi:2008jd,Nosaka:2017ohr}.)
This theory is the simplest example in a class of ${\cal N}=4$ linear quiver Chern-Simons matter theories, which are realized by taking the low energy limit of a UV field theory on D3-branes in a type IIB brane configuration. 
The type IIB brane configuration of the simplest linear quiver theory is given by Table.\ref{GWN4BraneSetup}.
\begin{table}[htbp]
\begin{center}
\begin{tabular}{|c|c|c|c|c|c|c|c|}
\hline
 & $012$ & $3$ & $456789$ \\ 
\hline 
NS5-brane & $\bigcirc $  &  $\times$ & [456]  \\ 
\hline
$N_1$ D3-branes & $\bigcirc $  & $\bigcirc $ &  $\times$ \\ 
\hline
$(1,k)$5-brane & $\bigcirc $  &  $\times$ & $[(47)_\theta(58)_\theta(69)_\theta]$  \\ 
\hline
$N_2$ D3-branes & $\bigcirc $  & $\bigcirc $ & $\times$  \\ 
\hline
NS5'-brane & $\bigcirc $  &  $\times$ & [456]  \\ 
\hline
\end{tabular} 
\caption{Type IIB Brane configuration describing the minimal linear quiver theory.
Here $\bigcirc$ indicates that the corresponding brane extends in all directions specified on its above column and $\times$ means that it does not in any. The numbers inside the bracket represent the directions of the extension, where $(ij)_\theta$ stands for a single direction in the $x^ix^j$-plane with the angle $\theta=\arctan k$ from the $x^{i}$ axis.  }
\label{GWN4BraneSetup}
\end{center}
\end{table} 
The field contents of such a UV field theory of the simplest model are the two $\cN=4$ vector multiplets associated with U$(N_1)\times \text{U}(N_2)$ and one hypermultiplet in the bi-fundamental representation in U$(N_1)\times\text{U}(N_2)$.
In the terminology of 3d ${\cal N}=2$ representations, here a 3d ${\cal N}=4$ vector multiplet consists of a vector multiplet plus a adjoint chiral multiplet, while a hypermultiplet consists of a pair of chiral multiplets.

These massless supersymmetric multiplets arise from open strings ending on the $N_1$ D3-branes, those on the $N_2$ D3-branes and the ones connecting the two stacks of D3-branes, respectively.
The $(1,k)$5-brane induces the `twisted' mass for the $\cN=4$ vector multiplets \cite{Kitao:1998mf,Bergman:1999na}: for the U$(N_1)$ vectormultiplet the ${\cal N}=2$ Chern-Simons term with level $k$ plus the cocmplex mass term proportional to $k$ for the adjoint chiral multiplet; for the U$(N_2)$ vectormultiplet the same terms with $k$ replaced with $-k$.
The twisted mass breaks SO$(4)_{\rm UV}$ R-symmetry to SO$(3)_{\rm UV}$. 

In the low energy limit this system gets strongly coupled and flow to the conformal fixed point, which is nothing but the minimal $\cN=4$ Chern-Simons theory mentioned above. 
In the IR limit, the $\text{SO}(3)_{\rm UV}$ R-symmetry is enhanced to SO$(4)_{\rm R}$, which can be explicitly seen by integrating out the massive vector multiplet and adjoint chiral multiplets except the Chern-Simons gauge field. 
In addition to the other global symmetry this system enjoys the parity invariance which exchanges the two gauge fields when $N_1=N_2$.
When the two ranks are different, the system is invariant under the exchange of the ranks as well as the levels.

\subsection{Moduli space} 
\label{modulispace}

In this subsection we briefly comment on the moduli space of the minimal $\text{U}(N_1)_k\times \text{U}(N_2)_{-k}$ $\cN=4$ Chern-Simons theory.
We will use the result of this analysis for a quick check of Seiberg-like duality discussed in the next subsection.
For convenience we write down the Lagrangian in the Euclidean space following the convention used in \cite{Yokoyama:2013pxa} ($\kappa=k/(4\pi)$)
\beal{
\cL
=& \tr \bigg[\kappa \Bigl(-\varepsilon^{\mu\nu\rho}\bigl(A_\mu\partial_\nu A_\rho +{2\over3}A_\mu A_\nu A_\rho\bigr)+ \varepsilon^{\mu\nu\rho}\bigl(\wt A_\mu\partial_\nu \wt A_\rho +{2\over3}\wt A_\mu \wt A_\nu \wt A_\rho\bigr) \Bigr)  +D_\mu  Q_A ^\dagger D^\mu Q^A -  {\psi^\dagger}^A \sla D \psi_A  \nn
&+ {1\over 2\kappa}(- {\psi^\dagger}^A \psi_A  Q^\dagger_C Q^C  +  \psi_A {\psi^\dagger}^A Q^B  Q^\dagger _B + \varepsilon_{CD} \varepsilon_{AB} {\psi^\dagger}^A  Q^C {\psi^\dagger}^B  Q^D +\varepsilon^{AB} \varepsilon^{CD} Q^\dagger_C  \psi_A Q^\dagger_D \psi_B )\bigg] +V_q.
\label{N4GWaction}
}
Here $A,B=1,2$ and the covariant derivative for $Q^A$ and $\psi_B$ is defined as $D_\mu X = \partial_\mu X +A_\mu X -X \wt A_\mu $.
$V_q$ is the scalar potential given by% 
\footnote{ 
This form of the scalar potential is obtained after rewriting the one given in \cite{Yokoyama:2013pxa} using an identity for SU(2) indices. 
}
\beal{ 
V_{q} =\tr [ T_C (T_C)^\dagger ]
}
where $T_C = {1\over 2\kappa} \varepsilon_{AB} Q^A Q^\dagger_C Q^B$. 

The vacuum moduli space is the solutions $(Q^1,Q^2)$ of $T_1=T_2=0$, modded by the U$(N_1)\times \text{U}(N_2)$ gauge transformations.
The generic point on the moduli spcae can be characterized as follows.
Due to the (generalized) parity invariance we can assume $N_1\le N_2$ without loss of generality.
By using the gauge degrees of freedom the complex scalar  $Q^1{}^a_{i}\; (1\leq a \leq N_1, 1\leq i \leq N_2 )$ can be diagonalized so that
\beal{  
Q^1{}^a_{ i} =  \left\{
\begin{array}{cl}
 r_a \delta^a_{ i} & (1 \leq  i \leq N_1) \\
 0 & (N_1 <  i \leq N_2) \\
\end{array}
\right.
}
where $r_a$ are real positive numbers.
The residual gauge symmetry is the diagonal $U(1)^{N_1}$. 
Now, in the generic situation where all $r_a$ are different with each other, the vacuum equation $T_C=0$ requires also $Q^2$ to be diagonal
\beal{  
Q^2{}^a_{ i} =  \left\{
\begin{array}{cl}
q_a \delta^a_{ i} & (1 \leq  i \leq N_1) \\
0 & (N_1 <  i \leq N_2) \\
\end{array}
\right.
}
with $q_a$ some complex numbers.

The diagonal U(1)$^{N_1}$ gauge symmetry fixes all the degrees of freedom of the gauge fields except its zero modes in the Cartan part, which span the extra directions in the moduli space since the abelian gauge field does not couple with the bi-fundamental matter fields.
The gauge flux quantization gives the $2\pi$ periodicity for the range of each zero mode, while the level $k$ Chern-Simons interaction breaks the U(1)$^{N_1}$ to $\mathbb{Z}_k^{N_1}$ (see \cite{Imamura:2008nn,Martelli:2008si} for example).
These zero modes become coordinates of the moduli space as $(S^1/\mathbb{Z}_k)^{N_1}$.

As a result, the classical moduli space is generically given by {$(\mathbb{C}^2/\mathbb{Z}_k)^{\text{min}(N_1,N_2)}/{\cal S}_{\text{min}(N_1,N_2)}$. We suspect that the moduli space does not receive any quantum correction and is classically exact at least for non-bad theories as in the cases with the Higgs phase in a linear quiver non-Chern-Simons theory and with $\cN=6$ ABJM theory. 
This is indeed supported from the superconformal index computed in section \ref{SCindex}.
We leave the proof thereof to a future work. In what follows, we confirm that this moduli space is consistent with the analysis below that there exists a decoupled sector of real dimension $4$ in the duality between pairs of non-bad theories.%
\footnote{ 
Precisely speaking the decoupled sector turns out to couple to the other sector through the topological current. 
}

\subsection{Comments on level-rank duality} 
\label{levelrankduality}

This minimal $\cN=4$ Chern-Simons theory is expected to enjoy the level-rank (or Seiberg-like) duality \cite{Giveon:2008zn}, that is, the duality between the $\text{U}(N_1)_k\times \text{U}(N_2)_{-k}$ theory and the $\text{U}(k-N_2)_k\times \text{U}(k-N_1)_{-k}$ theory.
The two theories are indeed related under the exchange of the two NS5-branes in the tpe IIB brane setup given in table \ref{GWN4BraneSetup} by taking into account the Hanany-Witten brane creation/annihilation \cite{Hanany:1996ie}.\footnote{
See \cite{Yokoyama:2013pxa} for the first scratch of evidence of this self-duality from the large $N$ thermal free energy.
}

{
On the other hand, there is a dicrepancy in the $S^3$ partition functions between the two theories \cite{Nosaka:2017ohr}; in some cases the partition function even diverges though it is finite in the other theory in the supposed dual.
By introducing the FI parameters which regularizes the divergences, we instead found that in any Hanany-Witten pair the ratio of the partition functions takes the same expression which depends on $k$, $N_1$, $N_2$ only through $k-N_1-N_2$.
}
This mismatch of the overall factor may be a signal of the existence of some decoupled sector. Indeed, it was pointed out that such decoupled sector appears if there exists a monopole operator whose dimension computed in the UV theory saturates or violates the unitarity bound in the case without Chern-Simons interaction \cite{Yaakov:2013fza,Hwang:2015wna}.
In the terminology used in \cite{Gaiotto:2008ak} such theories with (naively) unitarity violating monopole operators are called {\it bad} theories, and the  divergences discovered in \cite{Nosaka:2017ohr} suggests the following classification of the minimal ${\cal N}=4$ Chern-Simons theories:
\begin{itemize}
\item [] The minimal ${\cal N}=4$ $\text{U}(N_1)_k\times \text{U}(N_2)_{-k}$ Chern-Simons theory is respectively
\begin{align}
\text{good/ugly/bad if }
k-N_1-N_2 \text{ greater than/equal to/smaller than } -1.
\label{goodbaduglyfromZS3}
\end{align}
\end{itemize}
The main motivation of this paper is to investigate whether this classification is valid and what such decoupled sector in the minimal $\cN=4$ Chern-Simons theories is by using a superconformal index. 
 
The existence of the decoupled sector can be understood from the dimension of the moduli space.
For example, let us consider the minimal $\cN=4$ theory with $\text{U}(1)_3\times \text{U}(1)_{-3}$, whose moduli space is of real dimension $4$.
This theory is suggested to be dual to the one with $\text{U}(2)_3\times \text{U}(2)_{-3}$.
The dimension of the moduli space in the latter theory, however, is $8$ and does not match with the former theory.
We suspect that the discrepancy is explained as the presence of some decoupled sector in the latter theory, as in the cases without Chern-Simons terms.

In what follows, we study such decoupled sectors from a superconformal index, which encodes the BPS spectrum of the theory and is useful to analyze the system more precisely.

\section{Superconformal index}
\label{SCindex} 

In this section we examine the IR aspects of the ${\cal N}=4$ minimal $\text{U}(N_1)_k\times \text{U}(N_2)_{-k}$ superconformal Chern-Simons theory proposed in section \ref{levelrankduality} by studing the superconformal index.
After introducing the definition of the superconformal index in section \ref{generalexpression}, in section \ref{analytic} we first provide an analytic computation of the superconformal index.
Though this analytic computation works well only for the abelian case $N_1=N_2=1$, it and also tells us a technical obstacle associated with the bad theory.
In section \ref{perturbative} we adopt the small $x$ expansion.
Through an argument on its convergence we provide the good/bad classifiation for the theory, which coincides with the one suggested from the convergence of the $S^3$ partition function \eqref{goodbaduglyfromZS3}.
Finally we compare the superconformal indices of the two theories related under the Hanany-Witten transition in several examples.
As we have expected, we observe that their ratio takes the identical form as the superconformal index of a hypermulitplet.

\subsection{General expression} 
\label{generalexpression}

In this section we study the superconformal index of minimal $\cN=4$ $\text{U}(N_1)_k\times \text{U}(N_2)_{-k}$ Chern-Simons matter theories.
First let us define a superconformal index by introducing fugacities of the global symmetry which commutes with the chosen $\cN=2$ supersymmetry.
This requires a special care as we shall explain below. 

We compute the superconformal index by the supersymmetric localization so that 
we deform a minimal $\cN=4$ Chern-Simons theory to be an $\cN=3$ UV field theory described in section \ref{N4theory} in the free theory limit. 
We emphasize that the SO(4)$_\text{R}$ symmetry is broken to SO(3)$_{\text{UV}}$ under the deformation. 
Hence we shall turn on the fugacity only for this SO(3)$_{\rm UV}$ symmetry to define a superconformal index of the system, which we denote by $\sigma$. 
Furthermore, there is a conserved topological current for each gauge group associated with any node in the quiver diagram. 
We also turn on a fugacity for each topological current, which we denote by $y,z$ respectively for $\text{U}(N_1)$ and $\text{U}(N_2)$.
Taking these into account the superconformal index for this system is defined by the following trace over the spectrum of the theory:
\begin{align}
{\cal I}_{k,N_1,N_2}(x,\sigma,y,z)=\Tr[ (-1)^F e^{-\beta'\{Q,Q^\dagger\}} x^{D+J_3} \sigma^A y^{m_{\rm tot}} z^{n_{\rm tot}} ]
\label{IndexDefinition}
\end{align}
where $Q$ is a nilpotent supercharge and $Q^\dagger$ is its BPZ conjugate, while $D$, $J_3$, $A$, $(m_\text{tot},n_\text{tot})$ are respectively the dilatation, the angular momentum, the Cartan generator for SO$(3)_\text{UV}$ and the total magnetic charge of U$(N_1)\times \text{U}(N_2)$.
One can show that \eqref{IndexDefinition} is actually independent of $\beta'$ so that only BPS states, which satisfy $\{Q,Q^\dagger\}=D-J_3-R = 0$, contribute to the index.

To perform the supersymmetric localization, 
we first rewrite \eqref{IndexDefinition} as the path integral form over $\mbf S^2 \times \mbf S^1$.  
Then we take the weak coupling limit so that the evaluation of the path integral by the WKB approximation becomes exact. As a result each supersymmetric multiplet contributes to the index independently under the saddle points parametrized by magnetic charges and holonomies for U$(N_1) \times {\rm U}(N_2)$, which 
we denote by $(m_a,n_i)$ and $(u_a,v_i)$ with $a=1,\cdots, N_1, i=1,\cdots, N_2$, respectively. 
Then the 1-loop contribution of $\cN=4$ U$(N_1) \times {\rm U}(N_2)$ vector multiplet is given by
\beal{
{\cal I}_\text{vec}&=
\prod_{a\neq b}^{N_1}(1-x^{|m_a-m_b|}u_au_b^{-1})
\prod_{i\neq j}^{N_2}(1-x^{|n_i-n_j|}v_iv_j^{-1}), 
\label{Ivec}
}
and that of the bi-fundamental hypermultiplets is 
\beal{
{\cal I}_\text{hyp}&=
\prod_{a=1}^{N_1}
\prod_{j=1}^{N_2}
\frac{
(x^{|m_a-n_j|+\frac{3}{2}}\sigma^{-1}u_a^{-1}v_j;x^2)_\infty
}{
(x^{|m_a-n_j|+\frac{1}{2}}\sigma u_a v_j^{-1};x^2)_\infty
}
\frac{
(x^{|m_a-n_j|+\frac{3}{2}}\sigma u_a v_j^{-1};x^2)_\infty
}{
(x^{|m_a-n_j|+\frac{1}{2}}\sigma^{-1}u_a^{-1}v_j;x^2)_\infty
},
\label{Ihyp}
}
where we have used the Pochhammer symbol defined by 
\beal{
(z;x)_m :=& \prod_{n=0}^{m-1} (1-zx^n). 
}
Note that the contribution of the adjoint chiral multiplet in the $\cN=4$ vector multiplet completely cancels.
Including the vacuum and classical contribution the index is finally given by 
\begin{align}
{\cal I}_{k,N_1,N_2}(x,\sigma,y,z)
&=\frac{1}{N_1!N_2!}\sum_{\overrightarrow{m}\in \mathbb{Z}^{N_1} \atop \overrightarrow{n}\in\mathbb{Z}^{N_2}}
\int \prod_{a=1}^{N_1}\frac{du_a}{2\pi iu_a} 
\prod_{i=1}^{N_2}\frac{dv_i}{2\pi iv_i} y^{m_\text{tot}}z^{n_\text{tot}}
x^{\epsilon_0} u_a^{km_a} v_i^{-kn_i}
{\cal I}_\text{vec}
{\cal I}_\text{mat}
\end{align}
where $m_\text{tot}=\sum_{a=1}^{N_1}m_a$, $n_\text{tot}=\sum_{i=1}^{N_2}n_i$ and
\begin{align}
\epsilon_0&=\frac{1}{2}\sum_{a=1}^{N_1}\sum_{j=1}^{N_2}|m_a-n_j|
-\frac{1}{2}\sum_{a,b=1}^{N_1}|m_a-m_b|
-\frac{1}{2}\sum_{i,j=1}^{N_2}|n_i-n_j|. 
\end{align}
This expression may be obtained from the superconformal index of ABJM theory determined in \cite{Kim:2009wb} by excluding the contribution of one bifundamental hypermultiplet.   

The above expression of the index can be rewritten as a more compact form   
by using its invariance under the action of the Weyl group for ${\rm U}(N_1) \times {\rm U}(N_2)$:
\begin{align}
{\cal I}_{k,N_1,N_2}(x,\sigma,y,z)
&=\sum_{\overrightarrow{m}\in \mathbb{Z}^{N_1}/{\cal S}_{N_1} \atop \overrightarrow{n}\in\mathbb{Z}^{N_2}/{\cal S}_{N_2}}\frac{y^{m_\text{tot}}z^{n_\text{tot}}}{|W_{(\overrightarrow{m},\overrightarrow{n})}|}{\cal I}^{(\overrightarrow{m},\overrightarrow{n})}(x,\sigma)
\end{align}
where $\cS_N$ is the permutation group for $N$ elements and $|W_{(\overrightarrow{m},\overrightarrow{n})}|$ is the number of the permutations which fix the monopole configuration specified by $(\overrightarrow{m},\overrightarrow{n})$, and 
\begin{align}
{\cal I}^{(\overrightarrow{m},\overrightarrow{n})}(x,\sigma)=
x^{\epsilon_0}
\int
\prod_{a=1}^{N_1}\frac{du_a}{2\pi iu_a} u_a^{km_a}
\prod_{i=1}^{N_2}\frac{dv_i}{2\pi iv_i} v_i^{-kn_i}
{\cal I}_\text{vec}
{\cal I}_\text{mat}.
\end{align}

\subsection{Analytic computation in the abelian case}
\label{analytic}

Before considering the case with general ranks, let us start with the abelian case with $N_1=N_2=1$ with a general Chern-Simons coupling constant.
In this case we can compute the superconformal index analytically without difficulty.

In the minimal Gaiotto-Witten theory, the diagonal U(1) gauge field does not couple to the bi-fundamental matter, so that the corresponding holonomy integration can be trivially performed. 
In the abelian case, the only one non-trivial integration remains. 
We perform the remaining residue integration by deforming the integration contour to either the origin or infinity so as to avoid the poles arising due to the Chern-Simons interaction, as shown below. 

Let us first perform the integration of the diagonal U(1). This can be done by  changing integration variables such that 
\be 
w = \sigma^{-1} v_1/u_1.  
\ee
We fix the region of variables by $\sigma x^\half <1, x<1,\sigma x^{-\half} >1 $.
Then the integral form of the index becomes 
\beal{
{\cal I}_{k,1,1}^{(m_{\rm tot},n_{\rm tot})}(x,\sigma)
=& \oint {d{u} \over (2\pi i{u}) }u^{-k(m_{\rm tot} -n_{\rm tot})}  \oint {dw \over (2\pi iw) } \sigma^{km_{\rm tot}} w^{km_{\rm tot}} {(wx^{3/2} ; x^2)_\infty (w^{-1}x^{3/2}; x^2)_\infty  \over (w^{-1} x^{\half} ;x^2)_\infty (wx^{\half} ;x^2)_\infty  }, 
\notag
}
in which the integration of $u$ variable decouples from the other part. 
Performing this integration over $u$ gives  
\beal{
{\cal I}_{k,1,1}^{(m_{\rm tot},n_{\rm tot})}(x,\sigma)
=&\delta_{m_{\rm tot},n_{\rm tot}} \sigma^{km_{\rm tot}} \oint {dw \over (2\pi iw) } w^{km_{\rm tot}} {(wx^{3/2} ; x^2)_\infty (w^{-1}x^{3/2}; x^2)_\infty  \over (w^{-1} x^{\half} ;x^2)_\infty (wx^{\half} ;x^2)_\infty  }. 
}
It turns out that the contribution associated with the different monopole charges for two U(1) gauge groups vanishes.%
\footnote{ 
This statement may hold in the non-abelian case as well by considering the total monopole charges for U($N_1$) and U($N_2$). 
}
We perform the remaining integral by deforming the integration contour to either the origin or infinity so as to avoid the poles generated by the classical contribution, $w^{km_{\rm tot}}$ at the origin or infinity, which depends on the value of $m_{\rm tot}$.

When $m_{\rm tot} \geq  0$, the term $w^{km_{\rm tot}}$ is a pole around $w\sim\infty$. 
To avoid this pole, we deform the integration contour to the origin so that we pick up the poles inside the unit circle. 
\be 
{\cal I}_{k,1,1}^{(m_{\rm tot},n_{\rm tot})}(x,\sigma)
= \delta_{m_{\rm tot},n_{\rm tot}} \sigma^{km_{\rm tot}} \oint_{|w|<1} {dw \over (2\pi iw) } w^{km_{\rm tot}} {(wx^{3/2} ; x^2)_\infty (w^{-1}x^{3/2}; x^2)_\infty  \over (w^{-1} x^{\half} ;x^2)_\infty (wx^{\half} ;x^2)_\infty  }
\ee
We pick up poles inside the unit circle at $w= x^{2n+\half}$ with $n\geq0$, which come from the term $(w^{-1} x^{\half} ;x^2)_\infty$ in the denominator.\footnote{ 
As is the case without Chern-Simons interaction, we do not pick up the pole at the origin in this evaluation, which arises when $m_{\rm tot} = 0$ (see \cite{Krattenthaler:2011da} for example). This can be justified in the following way. Since the contribution coming from the pole at the origin is clearly ill-defined, one needs to regularize the contribution with $m_{\rm tot}=0$ for its evaluation. We regularize it by introducing a (discrete) background magnetic flux studied in \cite{Kapustin:2011jm} so that the pole at the origin disappears, which is turned off after the evaluation. Then the evaluation can be done as described in the main text.
}
Performing the residue integral we obtain 
\beal{
{\cal I}_{k,1,1}^{(m_{\rm tot},n_{\rm tot})}(x,\sigma)
=&\delta_{m_{\rm tot},n_{\rm tot}}\sigma^{km_{\rm tot}} \sum_{n\geq0} (x^{2n+\half})^{km_{\rm tot}} {((x^{2n+\half}) x^{3/2} ; x^2)_\infty ((x^{2n+\half})^{-1}x^{3/2}; x^2)_\infty  \over (x^{-2n};x^2)_n(x^2;x^2)_\infty ((x^{2n+\half})x^{\half} ;x^2)_\infty  } \nn
=& \delta_{m_{\rm tot},n_{\rm tot}}\sigma^{km_{\rm tot}} f^{m_{\rm tot}}(x)
}
where we set 
\be 
f^{m_{\rm tot}}(x)
= {1\over (x^2;x^2)_\infty} \sum_{n\geq0} (x^{2n+\half})^{km_{\rm tot}} {(x^{2n+2}; x^2)_\infty (x^{-2n+1} ; x^2)_\infty  \over (x^{-2n};x^2)_n (x^{2n+1}  ;x^2)_\infty  }. 
\ee

When $m_{\rm tot} < 0$, on the other hand, the term $w^{km_{\rm tot}}$ is a pole around $w\sim0$. To avoid this pole we deform the integration contour to the infinity so that we pick up the poles outside the unit circle. 
\be 
{\cal I}_{k,1,1}^{(m_{\rm tot},n_{\rm tot})}(x,\sigma)
= \delta_{m_{\rm tot},n_{\rm tot}} \sigma^{km_{\rm tot}} \oint_{|w|>1} {dw \over (2\pi iw) } w^{km_{\rm tot}} {(wx^{3/2} ; x^2)_\infty (w^{-1}x^{3/2}; x^2)_\infty  \over (w^{-1} x^{\half} ;x^2)_\infty (wx^{\half} ;x^2)_\infty  }.
\ee
Exchanging the integration variable such that $w \to \bar w = 1/w$, we find 
\beal{
{\cal I}_{k,1,1}^{(m_{\rm tot},n_{\rm tot})}(x,\sigma)=&\delta_{m_{\rm tot},n_{\rm tot}}\sigma^{km_{\rm tot}} \oint_{|\bar w|<1} {d\bar w \over (2\pi i\bar w) }\bar w^{-km_{\rm tot}} {(\bar w^{-1}x^{3/2} ; x^2)_\infty (\bar w x^{3/2}; x^2)_\infty  \over (\bar w x^{\half} ;x^2)_\infty (\bar w^{-1} x^{\half} ;x^2)_\infty  } 
}
which is the same as $\delta_{m_{\rm tot},n_{\rm tot}}\sigma^{km_{\rm tot}} f^{-m_{\rm tot}}(x)$.

As a result, we obtain  
\beal{
{\cal I}_{k,1,1}(x,\sigma, y,z)
=&\sum_{m_{\rm tot} \geq 0  } (yz)^{m_{\rm tot}} \sigma^{km_{\rm tot}} f^{m_{\rm tot}}(x)  + \sum_{m_{\rm tot} < 0  } (yz)^{m_{\rm tot}}  \sigma^{km_{\rm tot}} f^{-m_{\rm tot}}(x) \nn
=&\sum_{m_{\rm tot} > 0  } ( (yz\sigma^k)^{m_{\rm tot}} + (yz\sigma^k)^{-m_{\rm tot}} ) f^{m_{\rm tot}}(x) + f^{0}(x). 
} 
We have verified this expression by Taylor-expansion in terms of small $x$.%
\footnote{ 
To expand the superconformal index up to $x^{\nu_\text{th}}$ we truncate the summation over $m_{\rm tot}$, $n$ and the infinite product in the Pochhammer symbols as follows.
\begin{align}
\sum_{m_{\rm tot}=1}^\infty&\longrightarrow \sum_{m_{\rm tot}=1}^{m_\text{max}},\quad\quad\quad\quad \Bigl(m_\text{max}=\Bigl[\frac{2\nu_\text{th}}{k}\Bigr]\Bigr),\nonumber \\
\sum_{n=0}^\infty&\longrightarrow \sum_{n=0}^{n_\text{max}},\quad\quad\quad\quad \Bigl(n_\text{max}=\Bigl[\frac{1}{2km_{\rm tot}+1}\Bigl(\nu_\text{th}-\frac{km_{\rm tot}}{2}\Bigr)\Bigr]\Bigr),\nonumber \\
\frac{(x;x^2)_\infty}{(x^{1+2n};x^2)_\infty}
&\longrightarrow \frac{\prod_{j=0}^{\text{Floor}[\frac{x_\text{th}-n-1}{2}]}(1- x^{1+2j})
}{
\prod_{j=0}^{\text{Floor}[\frac{x_\text{th}-3n-1}{2}]}(1- x^{1+2n+2j})
}.
\end{align}
}

\subsubsection{Obstacle in bad theory for higher ranks}
In the case without Chern-Simons terms \cite{Hwang:2015wna} the benefit of such computation was that it works also for bad theories and enable us the direct comparison of the superconformal indices between the Hanany-Witten pairs.
In the same motivation below we show an attempt to generalize the above computation for the theories with higher ranks, though we end up with a difficulty for bad cases.

For simplicity we consider the case with $N_1=N_2=N$, which satisfy the s-rule bound $k\le N$.
First we consider the $u_a$-integrations, estimating the poles at the origin and at the infinity as (${\widetilde u}_a=u_a^{-1}$)
\begin{align}
{\cal I}\sim
\begin{cases}
\int du_a u_a^{km_a-(N-1)-1}=\int du_a u_a^{km_a-N}&\quad(u_a\sim 0)\\
\int du_a u_a^{km_a+(N-1)-1}=\int d{\widetilde u}_a e^{-km_a-N} &\quad(u_a\sim\infty)
\end{cases}
\end{align}
where in the middle $u_a^{\mp (N-1)}$ comes from ${\cal I}_\text{vec}$, which increases the singularity at $u_a=0,\infty$.

To see the difficulty we focus on the case $m_a>0$ and the contribution for the following choice of the poles: $u_a=(const)\cdot v_{i=a}$ which amounts to
\begin{align}
{\cal I}^{(\overrightarrow{m};\overrightarrow{n})}\sim \int \frac{dv_a}{2\pi iv_a} v_a^{-k(n_a-m_a)}{\cal I}_\text{vec}{\cal I}_\text{mat}.
\end{align}
Since the integrand is still a complicated function of $\{v_i\}$, we may want to perform the $v_i$-integration one by one by classifying $n_i$ in the same way as we did in $u_a$ integration.
Here we encounter a problem.
Due to the substitution $u_a=(const)\cdot v_a$, the $\text{U}(N)_k$ part of ${\cal I}_\text{vec}$ contributes in the same way as $\text{U}(N)_{-k}$ part.
Hence the estimation of poles at $v_i=0,\infty$ is modified as
\begin{align}
{\cal I}\sim 
\begin{cases}
\int dv_i v_i^{-k(n_i-m_i)-2(N-1)-1}=\int dv_i v_i^{-k(n_i-m_i)-2N+1} &\quad(v_i\sim 0)\\
\int dv_i v_i^{-k(n_i-m_i)+2(N-1)-1}=\int dv_i {\widetilde v}_i^{k(n_i-m_i)-2N+1} &\quad (v_i\sim \infty)
\end{cases},
\end{align}
which indicates:
for $n_i-m_i<-(2N-2)/k$ the integration has no pole at $v_i=0$ hence can be computed from the residues at the poles in $|v_i|<1$;
for $n_i-m_i>(2N-2)/k$ the integration has no pole at $v_i=\infty$ hence can be computed from the residues at the poles in $|v_i|>1$;
for $-(2N-2)/k\le n_i-m_i\le (2N-2)/k$ both poles are present with of order $N$ hence such computation does not work.
If $2N-2<k$, or the theory is good, the last case is satisfied only for $n_i=m_i$.
For $2N-2\ge k$, or the bad theory, however, there are several choices of the monopole charges where non-perturbative computation does not work.

\subsection{Perturbative computation}
\label{perturbative}

Now let us move on the case with general $N_1$, $N_2$.
As we argued above, analytic computation is not available for the ``bad'' theories with $k-N_1-N_2\le -2$.
The computation is already difficult, however, even for the good theories with higher ranks due to the increasing number of holonomy integrals.
In this section we instead consider the small $x$ expansion of the superconformal index from the beginning up to some finite order, which we shall call ``perturbative computation''.

For each $(\overrightarrow{m},\overrightarrow{n})$ we can compute ${\cal I}^{(\overrightarrow{m},\overrightarrow{n})}$ up to an arbitrary order of $x$, say $x^{\nu_\text{th}}$, in the following way:
(i) First expand the integrand ${\cal I}_\text{vec}{\cal I}_\text{hyp}$ around $x=0$; (ii) then expand each coefficient of small-$x$ expansion in the Laurent series of $(u_a,v_i)$; and (iii) finally pick up the monomial $\prod_au_a^{-km_a}\prod_iv_i^{kn_i}$ so that it compensate the Chern-Simons term $\prod_a u_a^{km_a} \prod_i v_i^{-kn_i}$.
This is indeed a procedure which is physically straightforward.
For each $(\overrightarrow{m};\overrightarrow{n})$ the bare monopole vaccum is not gauge invariant, which is indicated by the Chern-Simons term.
To obtain the gauge invariant states it requires additional excitations of the fields which are non-trivially charged under the gauge symmetry ((i) and (ii)) in an appropriate amount so that the net bound state is gauge invariant ((iii)).

To compute the full superconformal index we need to repeat these computations for all monopole charges.
In some cases the leading order of the non-zero contribution grows fast enough with respect to $|m_a|,|n_i|$, which allows us to compute the superconformal index up to arbitrary order $x^{\nu_\text{th}}$ by truncating the summation over the monopole charges at some finite $|m_a|,|n_i|$.

\subsubsection{Preliminary analysis of the classification \eqref{goodbaduglyfromZS3} }
\label{sec_singularity}

To see whether the last requirement is correct or not, let us estimate the leading order of $x$ for the following monoople charge $(m\ge 0)$
\begin{align}
(m_1,m_2,\cdots,m_{N_1};n_1,n_2,\cdot,n_{N_2})=(m,0,0,\cdots,0;m,0,0,\cdots,0).
\end{align}
For this choice the contribution from bare charge is
\begin{align}
x^{\epsilon_0}=x^{-\frac{m}{2}(N_1+N_2-2)}.
\end{align}
To cancel the gauge charge of the monopole $\prod_au_a^{km_a}\prod_iv_i^{-kn_i}=(u_1v_1^{-1})^{km}$, we also need to bring an appropriate number of monomials from each factors of ${\cal I}_\text{vec}{\cal I}_\text{hyp}$ \eqref{Ivec}, \eqref{Ihyp}.
Since we are interested in the leading order of $x$, here we adopt the choice with the smallest power of $x$.
In current choice of $(\overrightarrow{m};\overrightarrow{n})$, this is realized by picking $x^{\frac{1}{2}}\sigma^{-1}u_1^{-1}v_1$ from the second denominator in ${\cal I}_\text{hyp}$ $km$ times.
In total the leading order of $x$ in ${\cal I}^{(\overrightarrow{m};\overrightarrow{n})}$ is given as
\begin{align}
{\cal I}^{(m,0,0,\cdots,0;m,0,0,\cdots,0)}_{k,N_1,N_2}(x,\sigma)\sim x^{\frac{m}{2}(k-N_1-N_2+2)}.
\end{align}

From this result we conclude
\begin{enumerate}
\item If $k-N_1-N_2<-2$, the leading negative exponent of $x$ grows linearly in $m$ and not bounded from below.
Hence we cannot compute the small $x$ expansion of ${\cal I}$ in this approach.
\item If $k-N_1-N_2=-2$, the exponent of $x$ is bounded from below by $x^0$.
However, since infinitely many choices of monopole charges contribute to each order of $x$, it is again impossible to compute the small $x$ expansion of ${\cal I}$.
\item If $k-N_1-N_2>-2$, the expansion starts from a positive power of $x$.   
\end{enumerate}
Notice that the singularities for $k-N_1-N_2\le -2$ cannot be cured by introducing chemical potential for $m_\text{tot}=n_\text{tot}$.
To see this, let us work out similar power estimation for the following monopole charge
\begin{align}
(\overrightarrow{m};\overrightarrow{n})=(m+m_\text{tot},-m,0,\cdots,0;m+m_\text{tot},-m,0,\cdots,0),
\end{align}
with $m$ being an arbitrary integer, which contributes to the sector of $m_\text{tot}$.
For this monopole charge the contribution from bare monopole is $x^{\epsilon_0}=x^{-|2m+m_\text{tot}|-\frac{N_1+N_2-4}{2}(|m+m_\text{tot}|+|m|)}$.
The bare monopole is charged as $u_1^{k(m+m_\text{tot})} u_2^{-km} v_1^{-k(m+m_\text{tot})} v_2^{km}$ due to the Chern-Simons term.
Assuming $m>0$ and $|m|$ being large enought compared with $m_\text{tot}$ we find that the most economical way to cancel this charge and form a gauge singlet is to bring $(x^{\frac{1}{2}}u_1^{-1}v_1)^{k|m+m_\text{tot}|}\cdot (x^{\frac{1}{2}}u_2v_2^{-1})^{k|m|}$ from ${\cal I}_\text{hyp}$.
In total, the leading power of the superconformal index is
\begin{align}
{\cal I}_{k,N_1,N_2}^{(m+m_\text{tot},-m,0,0,\cdots,0;m+m_\text{tot},-m,0,0,\cdots,0)}(x,\sigma)\sim 
x^{\epsilon_0+\frac{k|m+m_\text{tot}|}{2}+\frac{k|m|}{2}}=x^{(k-N_1-N_2+2)(m+\frac{m_\text{tot}}{2})}.
\label{badforeachmtot}
\end{align}
Here we have used $m>0$ and $m\gg m_\text{tot}$ to reduce $|m+m_\text{tot}|=m+m_\text{tot}$, $|m|=m$, $|2m+m_\text{tot}|=2m+m_\text{tot}$.
The result \eqref{badforeachmtot} implies that for $k-N_1-N_2\le -2$ the non-analyticity of the superconformal index at $x=0$ appears already at each sector of $m_\text{tot}$.

This preliminary analysis implies that the classification \eqref{goodbaduglyfromZS3} done by using the convergence property of the three-sphere partition function works also for a superconformal index. 
In what follows we display the result of the perturbative computation of the superconformal index for several good or ugly theories.

\subsubsection{Results}
\label{perturbative_results}

We display the result of the perturbative computation of the superconformal index for various $k$, $N_1$, $N_2$.
We are especially interested in the pairs of $\text{U}(N_1)_k\times \text{U}(N_2)_{-k}$ theory and $\text{U}(k-N_2)_k\times \text{U}(k-N_1)_{-k}$ theory, which are suggested to be dual with each other from the Hanany-Witten transition.
Since $k-N_1-N_2=k-N_1-N_2$ transforms $k-N_1-N_2\rightarrow -k-N_1-N_2$ under the Hanany-Witten transition, such comparison is possible only for the pairs of $k-N_1-N_2=k-N_1-N_2=\pm 1$ because of the singularity for $k-N_1-N_2<-2$ argued in section \ref{levelrankduality}.\footnote{
For $k-N_1-N_2=0$ the pairs are different only in the sign of $k$.
In this case the superconformal indices trivially coincide in the dual pair.
}
For $|k-N_1-N_2|=1$ case, the explicit computation shows that the superconformal indices do not coincide in the proposed dual pair.
Nevertheless, their ratio simplifies and allows a physical interpretation as the contribution from an extra hypermultiplet.

We have computed the superconformal index for the proposed dual pairs with $N_1,N_2\le 2$, namely, $(k,N_1,N_2)=(1,0,0),(1,1,1),(2,0,1),(2,1,2),(3,0,2),(3,1,1),(3,1,3),(3,2,2)$ up to $x^5$.
Here for the cases with $N_1=0$ or $N_2=0$ the theory is the pure Chern-Simons theory and the superconformal index is trivially ${\cal I}(x,\sigma)=1$.
We have taken into account all the monopole charges in $|m_a|,|n_i|\le 20$, and found only small number of those in $|m_a|,|n_i|\le 10$ displayed in table \ref{contributingmn} contributes to the superconformal index, which supports this truncation is indeed exact up to $x^5$.
\begin{table}
\begin{tabular}{|c|c|c|l|}
\hline
$k$&$N_1$&$N_2$&$(\overrightarrow{m};\overrightarrow{n})$ (up to permutations and $(\overrightarrow{m},\overrightarrow{n})\rightarrow (-\overrightarrow{m},-\overrightarrow{n})$)\\ \hline\hline
$1$&$1$&$1$&$(0,0),(1,1),(2,2),(3,3),(4,4),(5,5),(6,6),(7,7),(8,8),(9,9),(10,10)$\\ \hline
$2$&$1$&$2$&$
(0;-1,1),
(0;0,0),
(1;0,1),
(2;0,2),
(3;0,3),
(4;0,4),
(5;0,5),
(6;0,6),
(7;0,7),
$\\
&&&$
(8;0,8),
(9;0,9),
(10;0,10)
$\\ \hline
$3$&$1$&$1$&$(0,0),(1,1),(2,2),(3,3)$\\ \hline
$3$&$2$&$2$&$(-5,5;-5,5),
(-4,4;-4,4),
(-3,3;-3,3),
(-2,2;-2,2),
(-1,1;-1,1),
$\\
&&&$
(0,0 ;0,0),
(-4,5;-4,5),
(-3,4;-3,4),
(-2,3;-2,3),
(-1,2;-1,2),
(0,1 ;0,1),
$\\
&&&$
(-4,6;-4,6),
(-3,5;-3,5),
(-2,4;-2,4),
(-1,3;-1,3),
(0,2 ;0,2),
(1,1 ;1,1),
$\\
&&&$
(-3,6;-3,6),
(-2,5;-2,5),
(-1,4;-1,4),
(0,3 ;0,3),
(1,2 ;1,2),
(-3,7;-3,7),
$\\
&&&$
(-2,6;-2,6),
(-1,5;-1,5),
(0,4 ;0,4),
(1,3 ;1,3),
(-2,7;-2,7),
(-1,6;-1,6),
$\\
&&&$
(0,5 ;0,5),
(1,4 ;1,4),
(-2,8;-2,8),
(-1,7;-1,7),
(0,6 ;0,6),
(1,5 ;1,5),
$\\
&&&$
(-1,8;-1,8),
(0,7 ;0,7),
(-1,9;-1,9),
(0,8 ;0,8),
(0,9 ;0,9),
(0,10;0,10)
$\\ \hline
\end{tabular}
\caption{
{
We list the monopole charges with non-vanishing contribution to the superconformal index up to ${\cal O}(x^5)$.
Here we have denoted only one element among each family generated by permutations and $(m_a;n_i)\rightarrow (-m_a;-n_i)$ whose contributions are identical (up to $(y,z,\sigma)\rightarrow (y^{-1},z^{-1},\sigma^{-1})$).
% $(\overrightarrow{m};\overrightarrow{n})\rightarrow (-\overrightarrow{m};-\overrightarrow{n})$
}
}
\label{contributingmn}
\end{table}

For our purpose of comparison between HW pairs it is convenient to express the superconformal index ${\cal I}$ in the letter index ${\widetilde {\cal I}}$ defined by the plethystic exponential
\begin{align}
{\cal I}(x,\sigma,y,z)=\PE[x,\sigma,y,z;{\widetilde {\cal I}}(x,\sigma,y,z)]=\exp[\sum_{n=1}^\infty \frac{1}{n}{\widetilde {\cal I}}(x^n,\sigma^n,y^n,z^n)],
\end{align}
with which the ratio is mapped to the difference
\begin{align}
\frac{{\cal I}}{{\cal I}'}=\PE[{\widetilde {\cal I}}-{\widetilde{\cal I}'}].
\end{align}
We have obtained the following letter indices
\begin{align}
{\widetilde {\cal I}}_{1,0,0}&=0,\nonumber \\
{\widetilde {\cal I}}_{1,1,1}&=(yz\sigma +y^{-1}z^{-1}\sigma^{-1})(x^{\frac{1}{2}}-x^{\frac{3}{2}}+x^{\frac{5}{2}}-x^{\frac{7}{2}}+x^{\frac{9}{2}})+{\cal O}(x^{\frac{11}{2}}),\nonumber \\
{\widetilde {\cal I}}_{2,0,1}&=0,\nonumber \\
{\widetilde {\cal I}}_{2,1,2}&=(yz\sigma^2+y^{-1}z^{-1}\sigma^{-2})(x^{\frac{1}{2}}-x^{\frac{3}{2}}+x^{\frac{5}{2}}-x^{\frac{7}{2}}+x^{\frac{9}{2}})+{\cal O}(x^{\frac{11}{2}}),\nonumber \\
{\widetilde {\cal I}}_{3,0,2}&=0,\nonumber \\
{\widetilde {\cal I}}_{3,1,1}&=x+(yz\sigma^3+y^{-1}z^{-1}\sigma^{-3})x^{\frac{3}{2}}-2x^2-(yz\sigma^3+y^{-1}z^{-1}\sigma^{-3})x^{\frac{5}{2}}+2x^3\nonumber \\
&\quad +2(yz\sigma^3+y^{-1}z^{-1}\sigma^{-3})x^{\frac{7}{2}}-3x^4-5(yz\sigma^3+y^{-1}z^{-1}\sigma^{-3})x^{\frac{9}{2}}+4x^5+{\cal O}(x^{\frac{11}{2}}),\nonumber \\
{\widetilde {\cal I}}_{3,1,3}&=(yz\sigma^3+y^{-1}z^{-1}\sigma^{-3})(x^{\frac{1}{2}}-x^{\frac{3}{2}}+x^{\frac{5}{2}}-x^{\frac{7}{2}}+x^{\frac{9}{2}})+{\cal O}(x^{\frac{11}{2}}),\nonumber \\
{\widetilde {\cal I}}_{3,2,2}&=(yz\sigma^3+y^{-1}z^{-1}\sigma^{-3})x^{\frac{1}{2}}
+x-2x^2
+2x^3
+(yz\sigma^3+y^{-1}z^{-1}\sigma^{-3})x^{\frac{7}{2}}
-3x^4\nonumber \\
&\quad -4(yz\sigma^3+y^{-1}z^{-1}\sigma^{-3})x^{\frac{9}{2}}
+4x^5
+{\cal O}(x^{\frac{11}{2}}).
\label{pertresults}
\end{align}

Interestingly, we find that the ratio of the superconformal index between Hanany-Witten pair is the same for all choices of $(k,N_1,N_2)$.
That is,
\begin{align}
\frac{{\cal I}_{k,N_1,N_2}}{{\cal I}_{k,k-N_2,k-N_1}}=\PE[x,\sigma,y,z;{\widetilde {\cal I}}_{(k,N_1,N_2)/(k,k-N_2,k-N_1)}] 
\end{align}
for $N_1>k-N_1$, with
\begin{align}
{\widetilde {\cal I}}_{(1,1,1)/(1,0,0)}=
{\widetilde {\cal I}}_{(2,1,2)/(2,0,1)}=
{\widetilde {\cal I}}_{(3,1,3)/(3,0,2)}=
{\widetilde {\cal I}}_{(3,2,2)/(3,1,1)}=(yz\sigma^k+y^{-1}z^{-1}\sigma^{-k})\frac{x^{\frac{1}{2}}-x^{\frac{3}{2}}}{1-x^2},
\label{ratioisfreehyper}
\end{align}
up to ${\cal O}(x^{\frac{11}{2}})$. 
As a result the indices of Hanany-Witten dual pairs turned out to coincide up to the contribution of one hypermultiplet up to the order. 
This result indicates that the Hanany-Witten duality holds up to a hypermultiplet. 
This implies that an ugly monopole operator decouples in the IR to form a hypermultiplet.%
\footnote{ 
Strictly speaking this ugly monopole operator does not totally decouple but couples to the other sector through the topological current. 
}

\section{Discussion}

In this paper we have considered the 3d ${\cal N}=4$ $\text{U}(N_1)_k\times \text{U}(N_2)_{-k}$ superconformal Chern-Simons theory coupled with a bifundamental hypermultiplet.
It was observed that the three-sphere partition function of this theory diverges if $k-N_1-N_2\le -2$ \cite{Nosaka:2017ohr}.
This suggests, following the argument \cite{Kapustin:2010mh,Yaakov:2013fza} in the case without Chern-Simons interactions, that the theory is {\it bad} when $k-N_1-N_2\le -2$ according to the good/ugly/bad classification in \cite{Gaiotto:2008ak}.
To check this classification we have studied the superconformal index of the theory and we have indeed found that there exists monopole operators with unitarity violating R-chage if $k-N_1-N_2\le -2$.

We have further computed the superconformal indices in small $x$ expansion for the pairs of theories with $k-N_1-N_2=\pm 1$ related by the Hanany-Witten transition.
As a result we have found that the superconformal indices of the two theories in pair coincide with each other up to an overall factor which is the same as the contribution of the hypermultiplet up to a certain order.
Notably, this is consistent with the dimension of moduli space $4\text{min}(N_1,N_2)$ of the theory: the difference between the dimensions the moduli space of the theories in a Hanany-Witten pair is $4|k-N_1-N_2|$, which coincides with the number of degrees of freedom of a hypermultiplet for $k-N_1-N_2=\pm 1$.
These results are again natural generalizations of what occurs in the case without Chern-Simons term.

It would be interesting to ask what happens in a pair with $|k-N_1-N_2|\ge 2$, that is, a pair of a good theory and a bad one.
From the dimension of the moduli space it is natural to expect that the number of decoupled hypermultiplets is $|k-N_1-N_2|$.
Unfortunately so far we do not have a method to compute the superconformal index of the bad theory.
The perturbative approach with the truncation of the summation over the monopole charges does not work for a bad theory.
In the case without Chern-Simons interactions the superconformal index is obtained in factorized form which is valid also for bad theories \cite{Hwang:2015wna}.
The computation, however, requires the theory to satisfy $|k|<|N_\text{f}-N_\text{a}|/2$ where $N_\text{f}$ and $N_\text{a}$ are the number of fundamental and anti-fundamental chiral multiplet in the UV field content (called as ``maximally chiral'') \cite{Benini:2013yva}.
This condition is not satisfied in our setup and the computation does not work due to the non-trivial poles at the origin and infinity.
It is desirable to establish an alternative technique to compute the superconformal index for our theory with $k-N_1-N_2\le -2$.

One possible approach is to introduce the fugacities for all the components of monopole charge $(m_1,m_2,\cdots,m_{N_1};n_1,n_2,\cdots,n_{N_2})$ not only for $m_\text{tot}=n_\text{tot}$.
This might remedy the singularity in the perturbative computation argued in section \ref{sec_singularity} and make the monopole summation convergent.
Once we obtain a resummed expression the original superconformal index will be obtained by sending all the fugacities to unity except for the one for $m_\text{tot}=n_\text{tot}$.

It would also be interesting to study the moduli space in more detail along the line of \cite{Nakajima:2015txa,Bullimore:2015lsa,Assel:2017jgo}.
Several generalizations of our setup could be studied in a similar manner.
We can also add an arbitrary number of fundamental hypermultiplets coupling with each gauge node.
As the extra hypermultiplets lift the R-charge of the monopole operators up, such generalizations are not only interesting by themselves but also can be easier than the original theory and would be helpful to understand the original theory.

\section*{Acknowledgement}

We would like to thank Chiung Hwang for helpful discussions and valuable comments.

\appendix

\bibliographystyle{utphys}
\bibliography{GWindexv2}

\end{document}